\RequirePackage{amsmath}
\documentclass[a4paper,fleqn,usenatbib]{mnras}

\usepackage{mathptmx}
\usepackage{txfonts}

\usepackage[T1]{fontenc}
\usepackage{ae,aecompl}

\usepackage{graphicx}	% Including figure files
\usepackage{amsmath}	% Advanced maths commands
\usepackage{amssymb}	% Extra maths symbols
\usepackage{subfig}
\usepackage{bm}
\usepackage{rotate}
\usepackage{rotating}
\usepackage{verbatim}
\usepackage{hyperref}
\newif\ifAMStwofonts
\AMStwofontstrue

\def\aj{\rmfamily{AJ~}}           % Astronomical Journal
\def\apj{\rmfamily{ApJ~}}         % Astrophysical Journal
\def\apjl{\rmfamily{ApJ~}}        % Astrophysical Journal, Letters
\def\apjs{\rmfamily{ApJS~}}       % Astrophysical Journal, Supplement

\def\aap{\rmfamily{A\&A~}}        % Astronomy and Astrophysics

\def\mnras{\rmfamily{MNRAS~}}     % Monthly Notices of the Royal Astron. Soc.
\def\pasj{\rmfamily{PASJ~}}       % Publications of the Astron. Soc. of Japan
\def\nat{\rmfamily{Nature~}}      % Nature
\def\ssr{\rmfamily{Space~Sci.~Rev.~}}% Space Science Reviews
\def\gs{\mathrel{\hbox{\rlap{\hbox{\lower4pt\hbox{$\sim$}}}\hbox{$>$}}}}
\def\ls{\mathrel{\hbox{\rlap{\hbox{\lower4pt\hbox{$\sim$}}}\hbox{$<$}}}}

\def\suzaku{{\it Suzaku~}}
\def\hst{{\it HST~}}

\def\swift{{\it Swift~}}
\def\xmm{{\it XMM-Newton~}}

\def\nustar{{\it NuSTAR~}}

\def\mrk335{{Mrk~335}}

\def\redchi{{\chi^2_\nu}}

\def\cstatdof{{C/{\rm dof}}}

\def\arcm{{\hbox{$^\prime$}}}
\def\arcs{{\hbox{$^{\prime\prime}$}}}

\def\obj{{III~Zw~2~}}
\title[ The changing reflector in \obj ]{ The changing source of X-ray reflection in the radio-intermediate Seyfert 1 galaxy \obj }

\author[A. G. Gonzalez et al.]{
A. G. Gonzalez\thanks{E-mail: agonzalez@ap.smu.ca},
S. G. H. Waddell,
L. C. Gallo
\\
Department of Astronomy and Physics, Saint Mary's University, 923 Robie Street, Halifax, NS, B3H 3C3, Canada\\
}

\date{Accepted 2017 December 4. Received 2017 November 17; in original form 2017 August 4}

\pubyear{2017}

\begin{document}
\label{firstpage}
\pagerange{\pageref{firstpage}--\pageref{lastpage}}
\maketitle

\begin{abstract}
We report on X-ray observations of the radio-intermediate, X-ray bright Seyfert 1 galaxy, III~Zw~2, obtained with \textit{XMM-Newton}, \textit{Suzaku}, and \textit{Swift} over the past 17-years. The source brightness varies significantly over yearly time scales, but more modestly over periods of days. Pointed observations with \xmm in 2000 and \suzaku in 2011 show spectral differences despite comparable X-ray fluxes. The \suzaku spectra are consistent with a power law continuum and a narrow Gaussian emission feature at $\sim6.4$ keV, whereas the earlier \xmm spectrum requires a broader Gaussian profile and soft-excess below $\sim2$ keV. A potential interpretation is that the primary power law emission, perhaps from a jet base, preferentially illuminates the inner accretion disc in 2000, but the distant torus in 2011. The interpretation could be consistent with the hypothesised precessing radio jet in \obj that may have originated from disc instabilities due to an ongoing merging event.
\end{abstract}


%%%%%%%%%%%%%%%%%%%%%%%%%%%%%%%%%%%%%%%%%%%%%%%%%%

%%%%%%%%%%%%%%%%% BODY OF PAPER %%%%%%%%%%%%%%%%%%%%%%

%######################%
\section{Introduction}
%######################%
\label{sect:introduction}
Active galactic nuclei (AGN) are responsible for some of the most energetic and luminous phenomena in the Universe. Emission from these objects spans a large range of the electromagnetic spectrum, from radio to $\gamma$-ray wavelengths. Since their identification by \cite{Seyfert1943} AGN have been classified into numerous categories based on their observed luminosity in these different energy bands. Through the analysis of X-ray emission from these objects we are able to study the innermost regions of AGN, exploring the central engine thought to be responsible for a vast number of different features such as relativistic jets and observed reflection spectra off the accretion disc. 

Of the various categories of AGN, a small set are radio-loud ($\sim10-20$ per cent) possessing large amounts of radio emission and typically displaying relativistic radio jets. The formation of these jets, however, is not well understood, though it is believed to be connected to accretion disc structure and mass accretion rate (e.g. \citealt{Merloni+2003,Falcke+2004,Kording+2006}). For example, \cite{Lohfink+2013} were able to determine that for 3C 120, a broad-line radio galaxy (BLRG), the innermost region of the accretion disc became destroyed via some disc instability and was thus launched as a jet. By studying sources that exhibit large amounts of both radio and X-ray emission, such as radio loud Seyfert galaxies, it is possible to probe this disc-jet connection through analysis of the disc and jet emission simultaneously (e.g. \citealt{Gallo+2006,Kataoka+2007,King+2017}). 

III Zw 2A (PG 0007+106, Mrk 1501,  $z=0.089$, hereafter referred to as III~Zw~2) is the brightest of a three member galaxy group. It is a peculiar object that has proven particularly difficult to classify in the standard AGN categories due to its highly variable radio band emission (e.g. \citealt{Aller+1985,Falcke+1999}) coupled with a Seyfert I nucleus (e.g. \citealt{Osterbrock1977}). Radio-loud AGN are typically housed in elliptical galaxies (e.g. \citealt{Kirhakos+1999}), though early studies of \obj indicate a spiral morphology (e.g. \citealt{HutchingsCampbell1983,Taylor+1996}). More recent studies using \textit{Hubble Space Telescope} (\hst) H-band images suggest an elliptical host galaxy \citep{Veilleux+2009}. A tidal bridge with knots of star formation exists between \obj and its nearest group neighbour III~Zw~2B \citep{Surace+2001} indicating an ongoing merger phase.

The radio emission from \obj is extremely variable, by as much as a factor of 20 \citep{Aller+1985} exhibiting a quasi-periodic activity cycle with period $P = 5.14\pm0.19$ yr \citep{Li+2010}. Variability is also present in the optical (e.g. \citealt{Lloyd1984,Clements+1995}) and X-ray (e.g. \citealt{KaastradeKorte1988,Salvi+2002}) bands. The extended radio emission, however, is weak (e.g. \citealt{Unger+1987,Brunthaler+2005}), which is expected from a Seyfert I galaxy. Notably, the radio jet present in \obj is the first superluminal jet discovered in a Seyfert galaxy \citep{Brunthaler+2000}. Exactly what causes the aforementioned immense radio variability is still a mystery. One explanation is that relativistic jets interacting with the interstellar medium (ISM) exciting molecular clouds near the central engine \citep{Falcke+1999}. Alternatively, the X-ray variability may be leading the radio by approximately 13 months \citep{Clements+1995}, suggesting changes in the inner disc region propagating outward altering the jet activity. 

\cite{Salvi+2002} studied the high-energy emission of \obj using archival data from 1977 up to and including the \xmm 2000 data. It was found that the X-ray spectrum was best described by a power law of slope $\Gamma \approx 1.7$ and an extremely broad Fe K$\alpha$ line (FWHM $\sim 140 000$ km s$^{-1}$) at 6.44 keV with no evidence for a soft excess or intrinsic absorption. No short term variability was detected in the X-ray light curves analysed by \cite{Salvi+2002}, however, long term (year time scales) 10-times flux variations were found alongside correlated X-ray and radio variability. They interpret the radio to optical emission as synchrotron radiation that is self-absorbed in the radio/millimetre energy range, with the X-ray emission being produced by inverse Compton scattering of UV seed photons from the accretion disc by electrons in the corona that also produce synchrotron radiation.

The broad Fe K$\alpha$ line found by \cite{Salvi+2002} has motivated this work to further analyse the spectral feature using data from the \suzaku and \swift satellites. Broad emission lines are thought to originate from segments of the accretion disc nearest the black hole (e.g. \citealt{Fabian+1989}). By studying the shape of these spectral features it is possible to obtain the emissivity profile (i.e. the pattern in which the disc is illuminated by the corona) (e.g. \citealt{WilkinsFabian2011}) and through analysis of the emissivity profile subsequently determine the geometry of the corona (e.g. \citealt{WilkinsFabian2012,Gonzalez+2017}) even if the emission is directed toward (e.g. \citealt{Wilkins+2014}) or away (e.g. \citealt{WilkinsGallo2015b}) from the disc. The detection of a broad feature in the X-ray spectrum \obj can therefore potentially be used to constrain the geometry of the corona, further motivating the analysis of recent observations.

In this work we study the X-ray spectrum of \obj using all available data since the 2000 \xmm observation (inclusive) in order to study the high-energy variability in an effort to explain its peculiar nature. The paper is organized in the following arrangement. First, we present the data used in this study and discuss the data reduction techniques performed on the raw data (Section \ref{sect:observations}). We then present our analysis on both the long and short term variability of the X-ray data (Section \ref{sect:variability}). Section \ref{sect:meanspectrum} displays the results of fitting the mean spectrum from each data set with simple phenomenological models before exploring more physically motivated spectral models. Finally, we discuss the results in greater detail (Section \ref{sect:discussion}) and conclude (Section \ref{sect:conclusions}).

%#####################################%
\section{Observations \& Data Reduction}
%#####################################% 
\label{sect:observations}
\begin{table*}
	\begin{center}
		\caption{X-ray observations of \obj used in this work. \xmm data corresponds to that taken by the EPIC pn detector. \suzaku include those taken by the two front-illuminated (FI) detectors, XIS0 and XIS3, and the HXD-PIN detector. Both XRT and BAT data from \swift were used and are shown below. Column (7) indicates the energy ranges used when performing broad band spectral fitting.} %\suzaku FI-XIS information is for a single instrument, not the combination of XIS0+XIS3.}
		\begin{tabular}{ccccccc}                
			\hline
			(1) & (2) & (3) & (4) & (5) & (6) & (7) \\
			Observatory & Observation ID & Start Date & Duration & Exposure & Counts & Energy Band \\
			&  & (yr.mm.dd) & (s) & (s) & & (keV) \\
			\hline
			\hline
			\xmm EPIC pn & 0127110201 & 2000.07.03 & 16311 & 10160 & 32835 & $0.3 -10$ \\
			\hline
			\suzaku FI-XIS & 706031010 & 2011.06.14 & 169080 & 81466 & 98392 & $0.7 - 10$ \\
			\suzaku HXD-PIN &  &  &  & 68980 & 16652 & $15 - 25$ \\
			\hline
			\swift XRT & 00036363001 & 2007.06.21 &  & 4828 & &  \\
			& 00036363002 & 2009.05.22 &  & 4284 & &  \\
			& 00036363003 & 2009.09.03 &  & 404 & &  \\
			& 00036363004 & 2010.02.01 &  & 75 & &  \\
			& 00036363005 & 2010.02.07 &  & 5701 & &  \\
			& 00036363006 & 2010.07.08 &  & 1128 & &  \\
			& 00036363007 & 2010.07.08 &  & 2316 & &  \\
			& 00036363008 & 2012.01.03 &  & 3284 & &  \\
			& 00049402001 & 2013.05.16 &  & 618 & &  \\
			& 00049402002 & 2016.11.07 &  & 172 & &  \\
			& 00049402003 & 2017.02.08 &  & 1209 & &  \\
			& 00049402004 & 2017.05.15 &  & 737 & &  \\
			& 00093000001 & 2017.05.18 &  & 718 & &  \\
			& 00093000002 & 2017.05.19 &  & 488 & &  \\
			& 00049402006 & 2017.07.29 &  & 792 & &  \\
			& Total & 2007$-$2017 &  & 26754 & 5361 & $0.5 - 7$ \\
			\swift BAT &  &  & $70$ months &  & 412 & $15-150$ \\
			\hline
			\label{tab:datalog}
		\end{tabular}
	\end{center}
\end{table*}

\begin{figure}
	\scalebox{1.0}{\includegraphics[width=\linewidth]{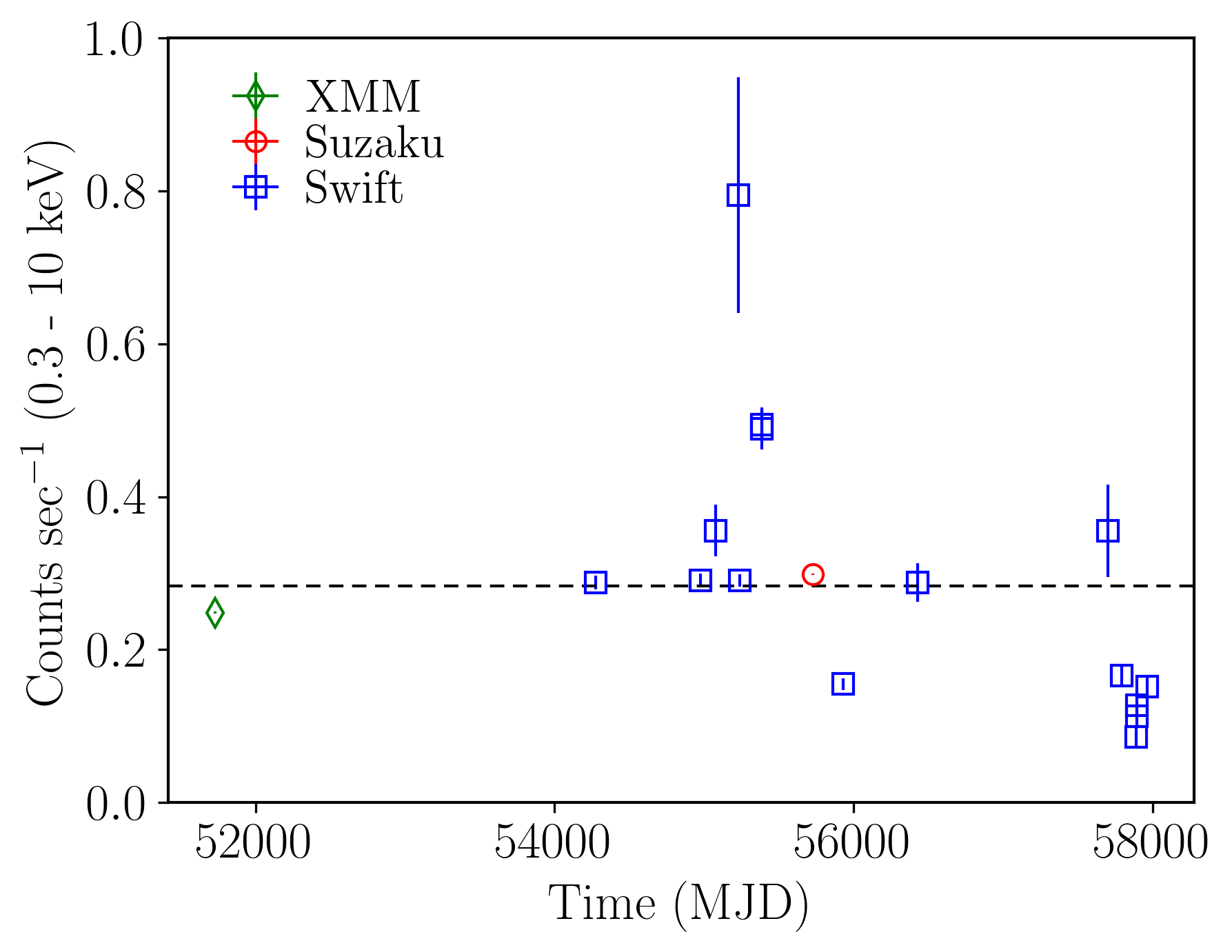}}
	\caption{The $0.3 - 10$ keV combined long term light curve of \obj with data from all three satellites expressed in \swift count rate. Error bars are included on all data points though they are not visible in all cases as the symbol is larger. The average total count rate is plotted as the dashed horizontal line.}
	\label{fig:longtermlightcurve}
\end{figure}

In this paper we use archival data obtained from the \xmm \citep{Jansen+2001}, \suzaku \citep{Mitsuda+2007}, and \swift \citep{Burrows+2005} satellites. Details of the observations are presented in Table \ref{tab:datalog}. 

\subsection{\xmm}
The EPIC pn \citep{Struder+2001} camera was operated in small window mode with the thin filter for the duration of the \obj observation. The \xmm Observation Data Files (ODFs) were processed using the \xmm Science Analysis System (\texttt{SAS v15.0.0}). EPIC response matrices were generated using the \texttt{SAS} tasks \texttt{ARFGEN} and \texttt{RMFGEN}. The light curve was extracted from the resulting event lists, with no significant background flaring detected. Source photons were extracted from a 35\arcs circular region centred on the source. Background photons were extracted from a 50\arcs circular off-source region. MOS data were examined by \cite{Salvi+2002} and determined to by consistent with those taken by the EPIC pn camera. We also determine that the MOS data is consistent with the EPIC pn data, though for simplicity here we use data from the EPIC pn camera exclusively.

\subsection{\suzaku}
\suzaku observed \obj in the XIS-nominal position, with the two front-illuminated CCDs (XIS0 and XIS3), the back-illuminated CCD (XIS1), and the HXD-PIN detectors. Cleaned event files from version 3 processed data were used to extract data products using \texttt{xselect}. For each XIS CCD, source photons were extracted from a 4\arcm circular region centred on the source. Background photons were extracted from a 3\arcm circular off-source region. Response files were generated using \texttt{xisrmfgen} and \texttt{xissimarfgen}. After determining that the XIS0 and XIS3 data were consistent the two were combined to create a single spectrum. XIS1 data were also found to be consistent with the FI detectors, though for simplicity here we use only the XIS0+XIS3 combined data. In accordance with the findings presented in \cite{Nowak+2011} we ignore the $1.88-1.92$ keV and $2.19-2.37$ keV ranges of the XIS0 and XIS3 spectra, and therefore of the combined spectrum, when performing spectral fits. 

An HXD-PIN spectrum was extracted following standard procedures.  The non-X-ray background (NXB) file corresponding to the observations was used to determine the common good-time interval between the NXB and the source data.  The resulting PIN exposure was $68980$ sec once corrected for detector dead time.  The cosmic X-ray background (CXB) was modelled using the provided flat response files and the PIN background spectrum was created by merging the CXB and NXB background files.  Examination of the PIN data yield a $11$ per cent detection of the AGN between $15-25$ keV.

\subsection{\swift}
Fifteen observations of \obj were obtained with the Swift XRT in window timing mode between 2007 and 2017 (Table \ref{tab:datalog}). Light curves and spectral data products were extracted with the online Swift-XRT data products generator\footnote{\href{http://www.swift.ac.uk/user\_objects/}{http://www.swift.ac.uk/user\_objects/}} \citep{Evans+2009}, which produces calibrated and background subtracted products.

The average exposure during each observation was $1784$ sec, but individual exposures ranged from $74$ sec to $5700$ sec.  Given the small counts in each exposure, the spectrum was generated from the combined fifteen observations over the ten-year span. The total combined exposure is $26754$ sec and the source is detected between $0.5-7$ keV.

Also included in the analysis is the Swift-BAT spectrum that was generated from the 70-month survey\footnote{\href{https://swift.gsfc.nasa.gov/results/bs70mon/}{https://swift.gsfc.nasa.gov/results/bs70mon/}} \citep{Baumgartner+2013}.

\subsection{General}
All spectral fitting was done using \texttt{XSPEC v12.9.1} \citep{Arnaud1996}. \xmm EPIC pn and \suzaku XIS spectra were binned in accordance with the optimal binning technique described in \cite{KaastraBleeker2016} and implemented in the \texttt{Python} code made available by C. Ferrigno\footnote{\href{https://cms.unige.ch/isdc/ferrigno/developed-code/}{https://cms.unige.ch/isdc/ferrigno/developed-code/}}. The average \swift source spectrum was grouped such that each bin contains a minimum of $20$ counts. In the plots presented the spectra have been re-binned for display purposes. Model likelihoods were evaluated using the \cite{Cash1979} statistic, modified in \texttt{XSPEC} as the $C$-statistic. Note that only the BAT data are Gaussian\footnote{\href{https://swift.gsfc.nasa.gov/analysis/threads/batspectrumthread.html}{https://swift.gsfc.nasa.gov/analysis/threads/batspectrumthread}}, however, $C$-statistics are presented here as well to simplify presentation. Parameters reported in the fits are in the rest frame of the source and errors on them correspond to a 90 per cent confidence level. The Galactic column density toward \obj of $5.39\times10^{20}$ cm$^{-2}$ \citep{Kalberla2005} is kept fixed in all spectral fits. The reported fluxes in the relevant models are produced using the \texttt{XSPEC} model \texttt{cflux}. Since the \suzaku data were taken in XIS-nominal position the HXD-PIN and XIS cross-normalisation is fixed at 1.16. Henceforth, references to \xmm data correspond to that taken by the EPIC pn camera, \suzaku to the combined XIS0+XIS3 data, and \swift to the XRT data. The PIN data are linked to the \suzaku data in all relevant fits, and similarly the BAT data are linked to the \swift data.

%#######################%
\section{Variability}
%#######################%
\label{sect:variability}

\subsection{Long Term}
\label{sect:longterm}
\begin{figure}
	\scalebox{1.0}{\includegraphics[width=\linewidth]{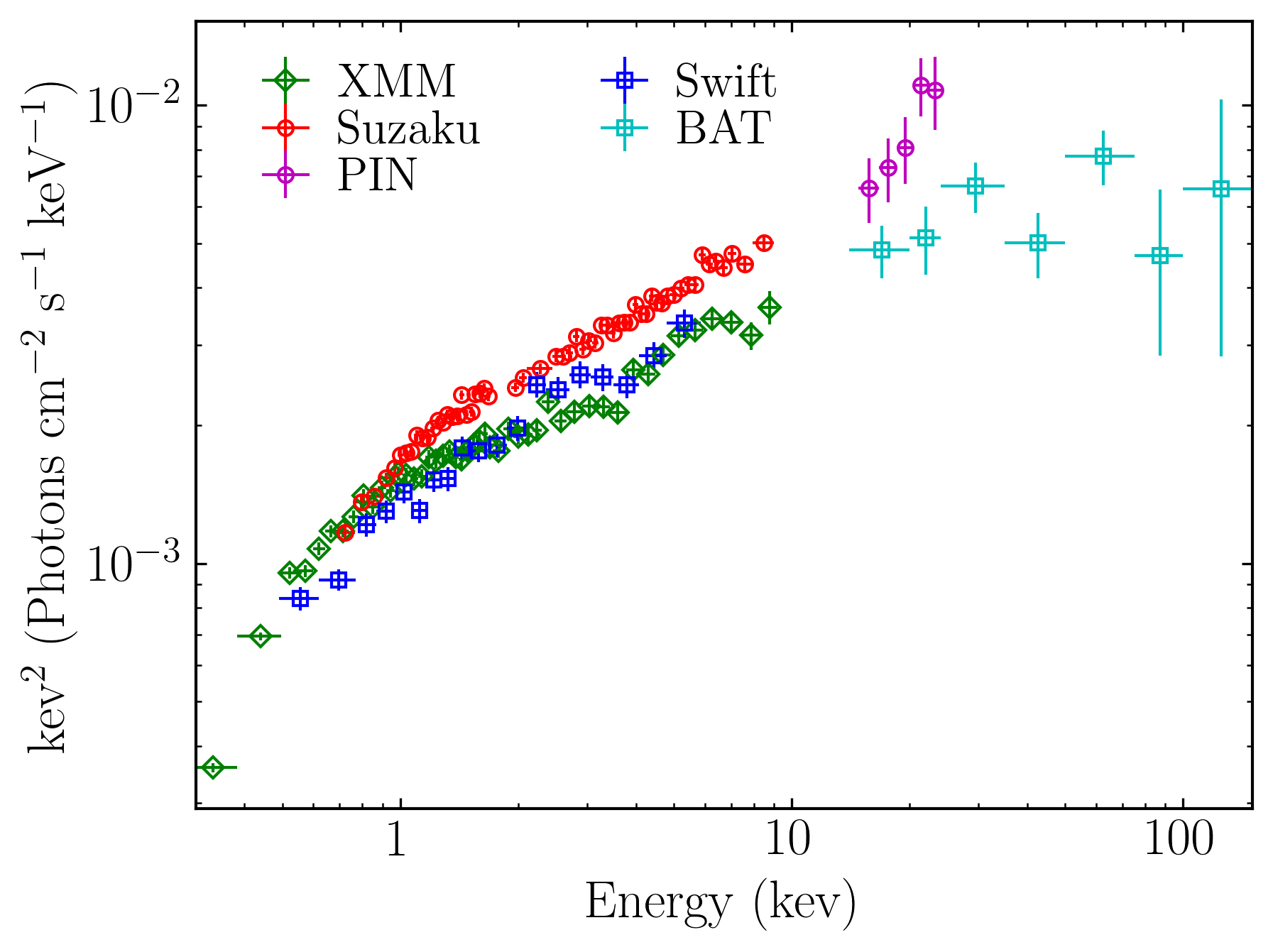}}
	\caption{The unfolded spectra of \obj for all the data fit by a power law with slope $\Gamma = 0$ to examine the change in spectral shape over time. No significant spectral changes are observed.}
	\label{fig:po0}
\end{figure}
Our analysis of \obj begins with examining its variability over the observations between July 2000 and July 2017 (Table \ref{tab:datalog}). Figure \ref{fig:longtermlightcurve} shows the combined long term $0.3-10$ keV light curve, which displays significant changes in count rate over the 17$-$year time span. Fitting the weighted mean count rate to the combined light curve ($0.283\pm0.001$ counts s$^{-1}$) results in a fit statistic of $\redchi = 35.47$. The fractional variability \citep{Edelson+2002} is $F_{\mathrm{var}} = 25\pm15$ per cent, further indicating the presence of long term X-ray variability in III~Zw~2. 

In the \swift light curve the average count rate ($0.228\pm0.003$ counts s$^{-1}$) produces a poor fit to the data ($\redchi = 61.45$) and has a fractional variability of $F_{\mathrm{var}} = 86\pm17$ per cent. Due to the sparse data, however, these significant deviations from the average count rate and variability cannot be further explored. The mean spectrum of these \swift observations is therefore used for the remainder of the spectral fitting and analysis. 

It is then necessary to examine spectral changes between the different observations, as shown in Figure \ref{fig:po0}. By unfolding the spectra against a power law model with slope $\Gamma = 0$ we are able to correct each spectrum for the effective area of its respective detector, allowing for a direct spectral comparison across different time epochs and instruments. The three data sets appear to be in approximate agreement with each other. The \suzaku data exhibit a higher flux with similar slope to both the \xmm and \swift data. The colour and marker scheme presented in Figure \ref{fig:po0} (\xmm - green diamond, \suzaku - red circle, PIN - magenta circle, \swift - blue square, and BAT - cyan square) is maintained throughout this work. 

\subsection{Short Term}
\label{sect:shortterm}
\begin{figure}
	\scalebox{1.0}{\includegraphics[width=\linewidth]{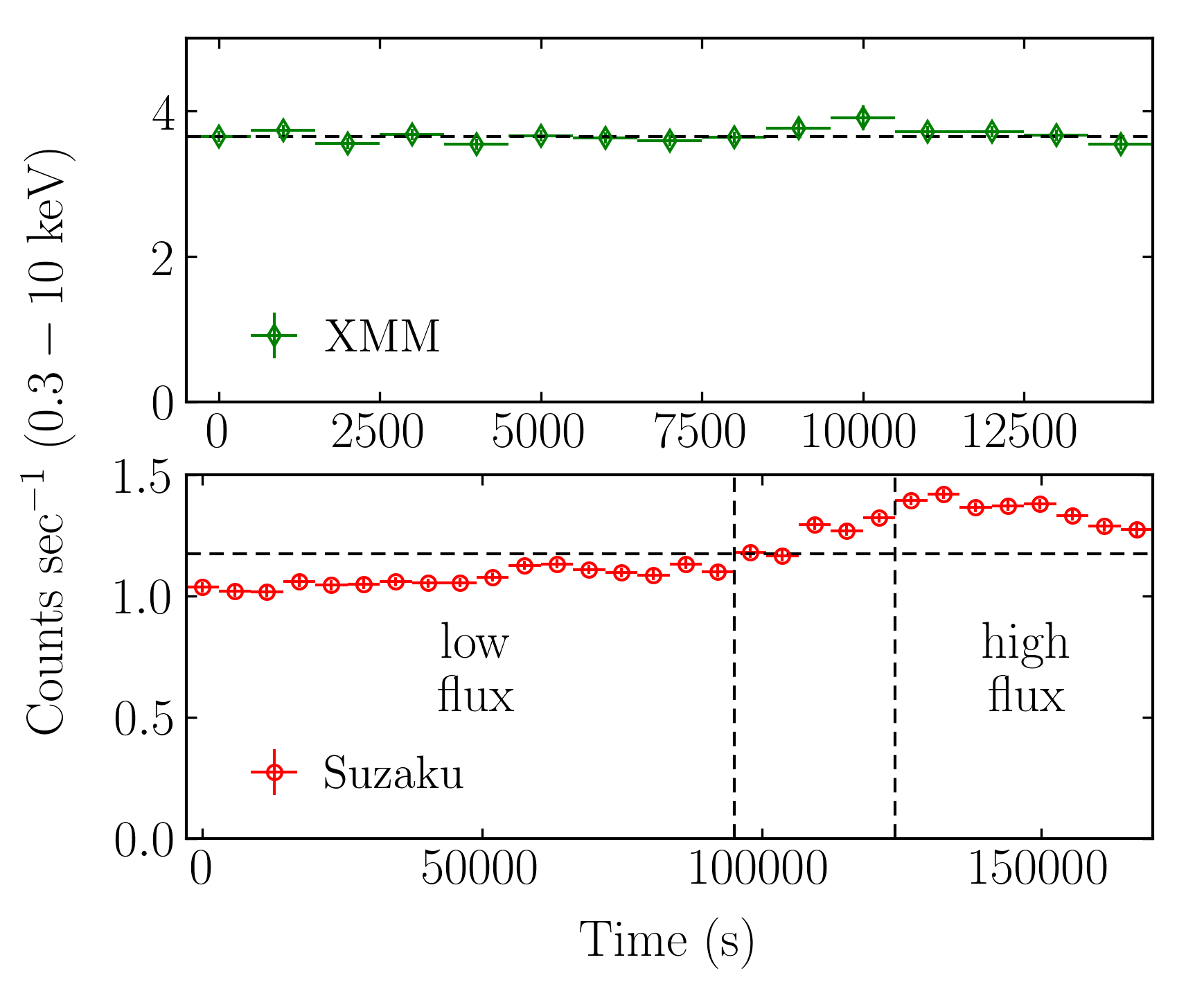}}
	\caption{The $0.3 - 10$ keV short term light curves of \obj for the \xmm (top panel) and \suzaku (bottom panel) observations. The \xmm data have been binned in $1000$ second bins and the \suzaku data have been binned according to the $96$ minute orbital period of the satellite. Vertical lines in the bottom panel denote the low (left of first line) and high (right of second line) flux states of the \suzaku observation. Horizontal dashes lines represent the average count rates in each panel.}
	\label{fig:shorttermlightcurve}
\end{figure}

The short term light curves for the \xmm and \suzaku observations are shown in Figure \ref{fig:shorttermlightcurve}. The \xmm light curve is flat and featureless throughout. Calculating the fractional variability ($F_{\mathrm{var}} < 2.3$ per cent) and fitting the mean ($3.65\pm0.02$ counts s$^{-1}$) to the light curve ($\redchi = 0.85$) indicate no significant variability during this observation, requiring no further investigation. With this it is therefore acceptable to use the mean spectrum of the \xmm data for the remainder of the spectral fitting and analysis.

\begin{figure}
	\scalebox{1.0}{\includegraphics[width=\linewidth]{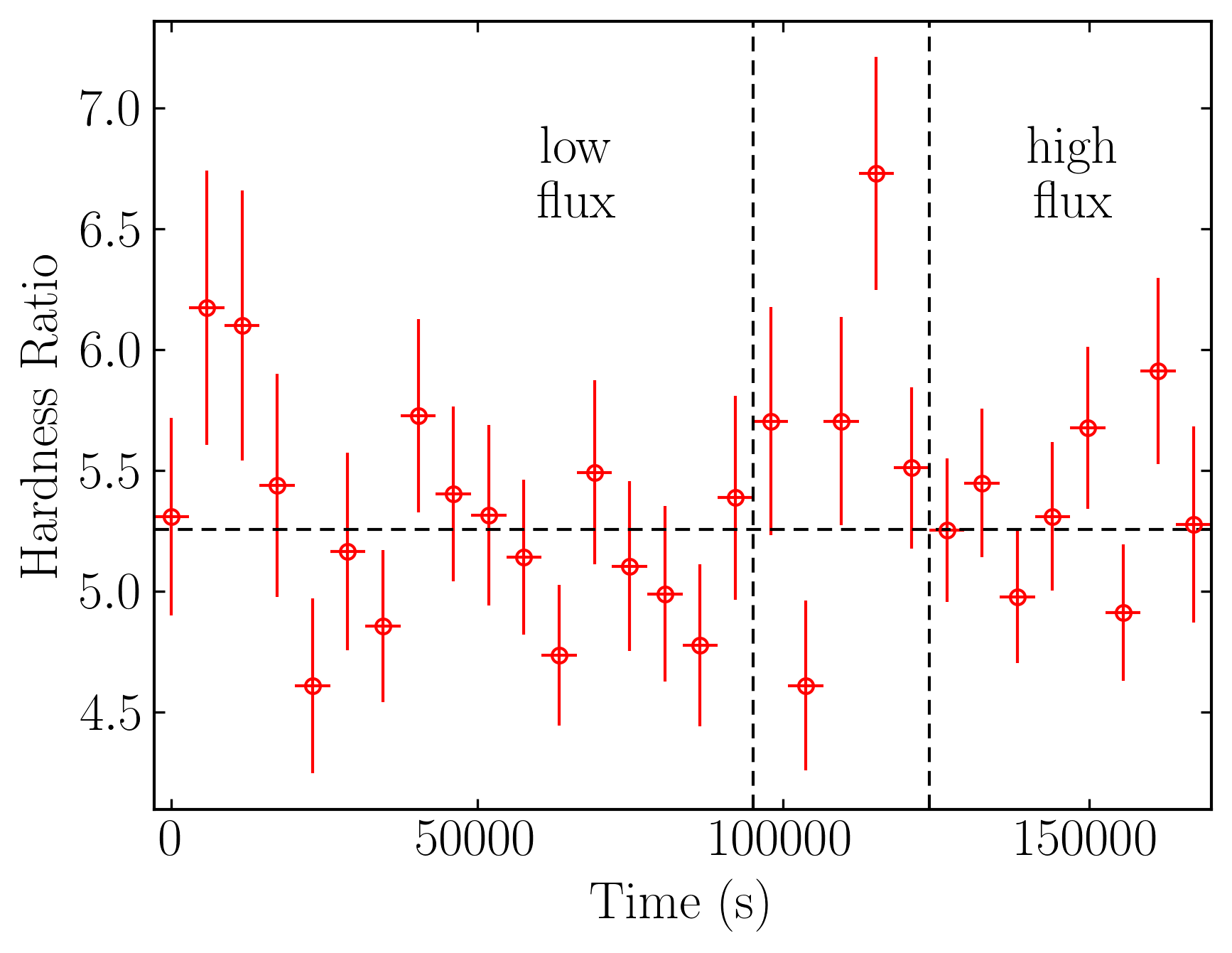}}
	\caption{Hardness ratio (hard/soft) between the $0.5 - 1$ keV (soft) and $2 - 10$ keV (hard) energy bands of the \suzaku observation of III~Zw~2. The dashed horizontal line represents the average ratio with vertical lines denoting the low (left of first line) and high (right of second line) flux states.}
	\label{fig:hardnessratio}
\end{figure}

The \suzaku observation is relatively flat for the first $\sim95$ ks before rising slightly for the remainder of the observation ($\sim75$ ks). Computing the fractional variability ($F_{\mathrm{var}} = 12\pm2$ per cent) and fitting the mean ($1.174\pm0.004$ counts s$^{-1}$) to the light curve ($\redchi = 40.78$) state the significance of this short term variability. Considering the small increase in count rate we examine the hardness ratio between the $0.5 - 1$ keV and $2 - 10$ keV bands of the \suzaku data, shown in Figure \ref{fig:hardnessratio}. Fitting the hardness ratio to the mean ($5.26\pm0.07$) finds moderate deviation ($\redchi = 1.41$). No corresponding step in the hardness ratio is observed at $\sim95$ ks as in the light curve.

\begin{figure}
	\scalebox{1.0}{\includegraphics[width=\linewidth]{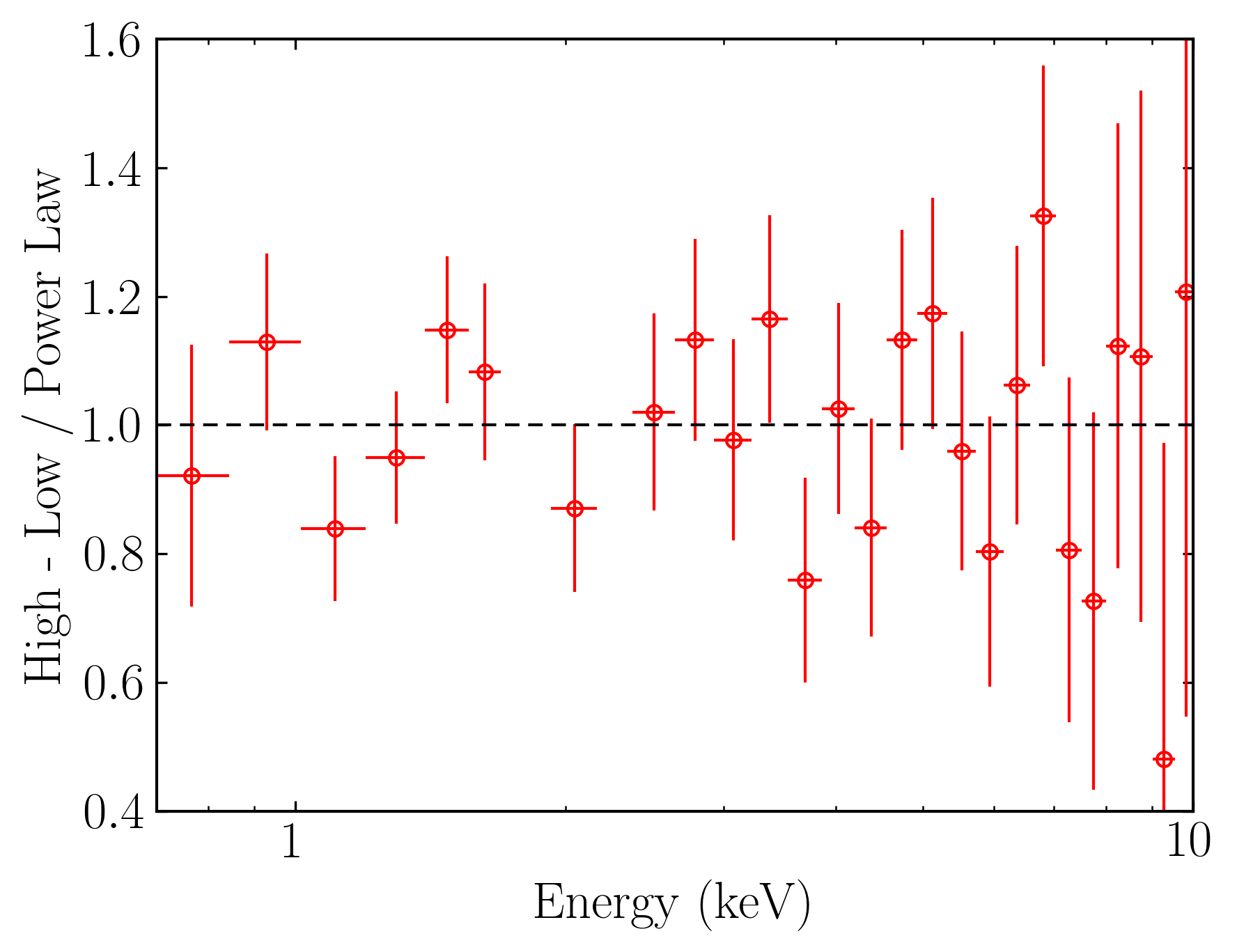}}
	\caption{Spectral comparison of the high and low flux states for \obj during the \suzaku observation. Taking the difference between the high and low flux states and fitting with a power law of slope $\Gamma = 1.53$ finds no significant spectral differences between them, supporting analysis of only the mean spectrum for the \suzaku observation.}
	\label{fig:hilocomp}
\end{figure}

The low flux state was taken to be the first $95$ ks of the observation and the high flux state was taken to be all data after $124$ ks. The resulting spectral difference (high-low) compared to a power law ($\Gamma = 1.53$) is shown in Figure \ref{fig:hilocomp}. Here the spectra have been re-binned to more clearly show any trends. There are no significant differences between the two flux states of the \suzaku data. Therefore, the mean spectrum of the \suzaku data is used for the remainder of the spectral fitting and analysis.

%#############################%
\section{Fitting the Mean Spectrum}
%#############################%
\label{sect:meanspectrum}

\subsection{High Energy X-ray Spectrum}
\label{sect:highenergy}
\begin{figure}
	\subfloat[\label{fig:2-10res}]{\includegraphics[width=\linewidth]{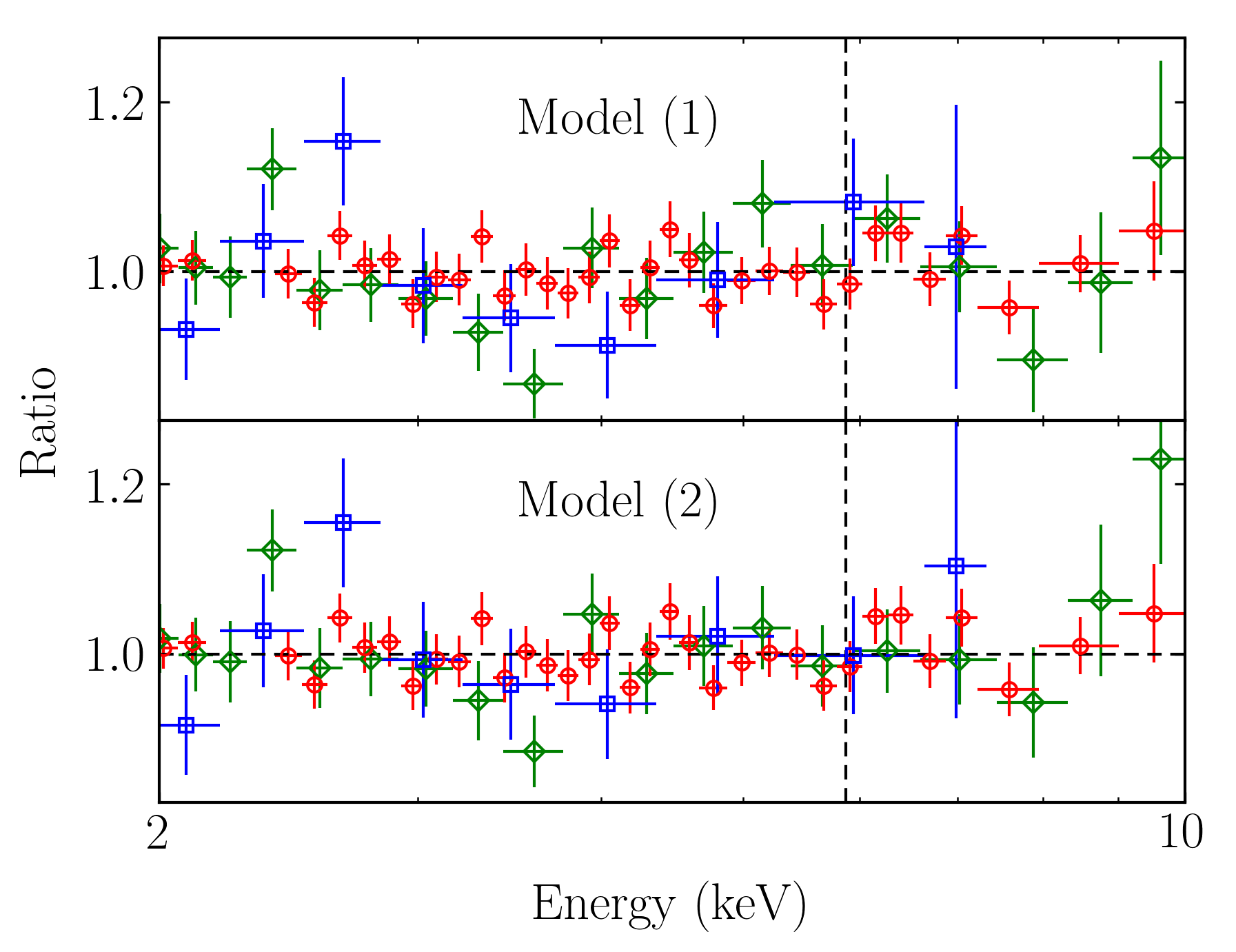}}
	\\
	\subfloat[\label{fig:ironline}]{\includegraphics[width=\linewidth]{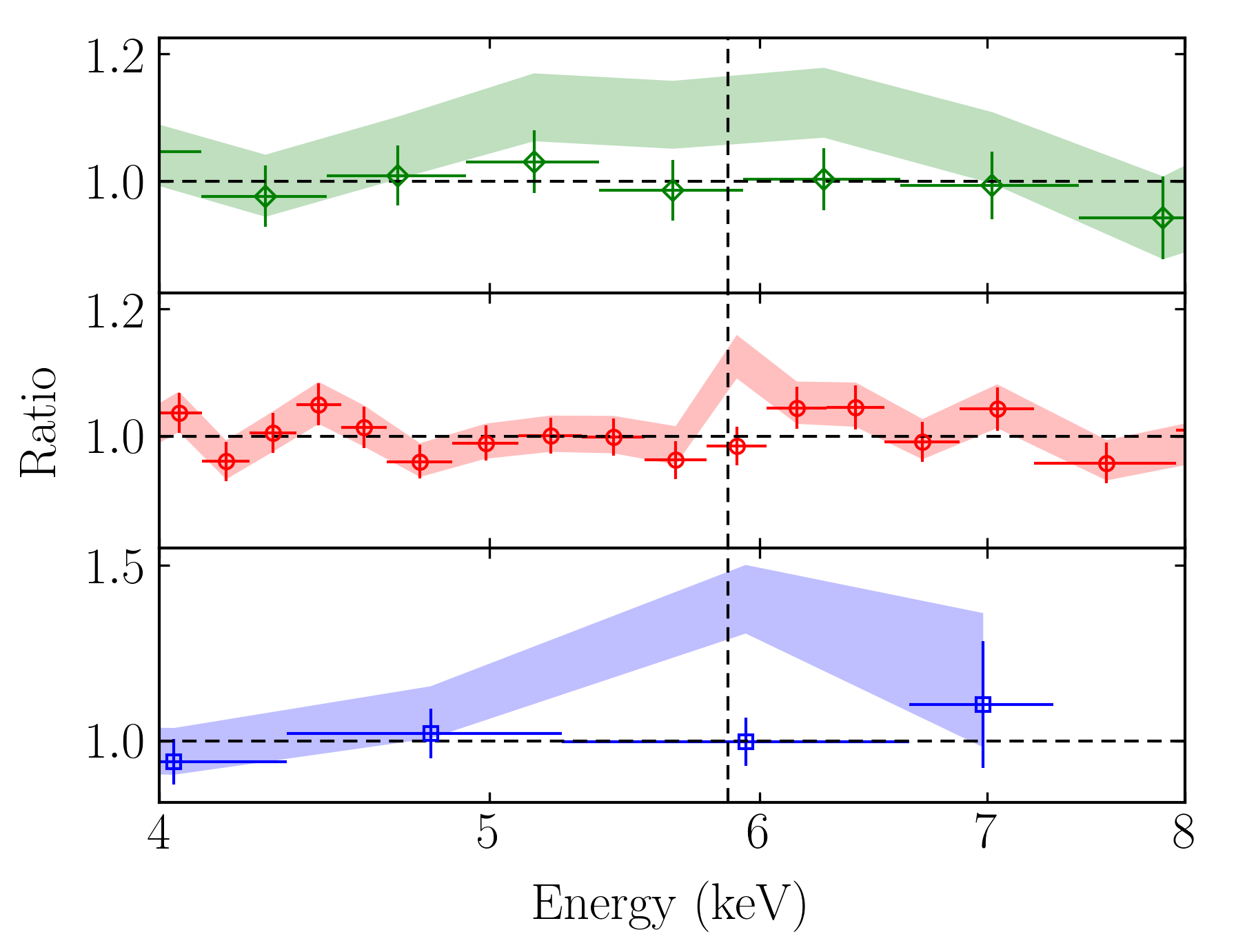}}
	
	\caption{(a) The ratio residuals produced by Model (1), a cut-off power law and a narrow Gaussian centred at $6.4$ keV, and Model (2), a cut-off power law and a broad Gaussian, as presented in Table \ref{tab:2-10params} fit over the $2-10$ keV energy range. (b) A close-up of the $4-8$ keV energy range for the three data sets. Shaded regions denote the best fit of a cut-off power law to the $2-10$ keV energy range excluding $4-7$ keV. Empty markers denote the fit given by Model (2). Vertical lines in both (a) \& (b) denote the rest frame Fe K$\alpha$ line energy of $6.4$ keV.}
\end{figure}

%%% USING THE NEW RBN SPECTRA
\begin{table*}
	\begin{center}
		\caption{The best fit parameters found by fitting the $2-10$ keV band of \obj for the three observations. In Model (1) the line energy of the Gaussian features is kept frozen at $6.4$ keV and the width is kept frozen at $1$ eV. In Model (2) both aforementioned parameters are left free to vary. The normalisation of the power law component is in units of photons keV$^{-1}$ cm$^{-2}$ s$^{-1}$ at $1$ keV.}
		%\textbf{RBN SPECTRA C-STAT}
		\begin{tabular}{cccccccc}                
			\hline
			Model & Data Set & $\Gamma$ & Norm$_{\mathrm{pl}}$ & $E_{\mathrm{K}\alpha}$ & $\sigma_{\mathrm{K}\alpha}$ & $EW_{\mathrm{K}\alpha}$ & $\cstatdof$ \\
			& & & ($\times10^{-3}$) & (keV) & (keV) & (eV) & \\
			\hline
			\hline
			
			& \xmm & $1.55\pm0.04$ & $1.42^{+0.09}_{-0.08}$ & & & $42^{+47}_{-34}$ & 103.40/102 \\
			(1) cpl+zgN & \suzaku & $1.54\pm0.02$ & $1.90\pm0.06$ & $6.4$ & $0.001$ & $39^{+17}_{-16}$  & 95.99/98 \\
			& \swift & $1.71\pm0.12$ & $1.88^{+0.30}_{-0.26}$ & & & $219^{+103}_{-124}$ & 93.45/79  \\
			
			\hline
			
			& \xmm & $1.61^{+0.07}_{-0.06}$ & $1.49^{+0.11}_{-0.10}$ & $6.26\pm0.55$ & $0.96^{+0.71}_{-0.58}$ & $312^{+239}_{-312}$ & 98.60/100 \\
			(2) cpl+zgB & \suzaku & $1.54\pm0.02$ & $1.90\pm0.06$ & $6.40^{+0.09}_{-0.05}$ & $<0.158$ & $39^{+17}_{-13}$  & 95.98/96 \\
			& \swift & $1.79^{+0.14}_{-0.13}$ & $2.03^{+0.35}_{-0.30}$ & $6.30^{+0.16}_{-0.13}$ & $0.22^{+0.21}_{-0.11}$ & $467^{+299}_{-262}$ & 85.70/77 \\
			
			\hline
			\label{tab:2-10params}
		\end{tabular}
	\end{center}
\end{table*}

We begin our spectral analysis of \obj by studying the $2-10$ keV energy range for the mean spectrum of each observation. To start we fit a simple cut-off power law (denoted cpl) and narrow Gaussian (denoted zgN) to each data set, with the power law cut-off energy frozen at $300$ keV (this value is maintained throughout the work presented as it was not found to improve the fit when left free to vary) and the Gaussian frozen at $E_{\mathrm{K}\alpha} = 6.4$ keV and width frozen at $\sigma_{\mathrm{K}\alpha} = 1$ eV. Good statistical fits to the \xmm and \suzaku data were found, with the \swift data having the poorest fit quality. The parameter values are presented as Model (1) in Table \ref{tab:2-10params} and the ratio residuals are displayed in the top panel of Figure \ref{fig:2-10res}. Though the \xmm and \suzaku fits produce $\Gamma$ values more closely in agreement with one another, the large error on the \swift fit does mean that all three values are in agreement. 

Freeing the line energy and width to allow fitting of a broad feature (denoted zgB) significantly improves the \xmm and \swift fits, indicating the presence of a broad spectral feature at $\sim6.3$ keV in these data. The \suzaku fit is unaffected by the free energy and breadth of the line and thus indicates a narrow feature at $6.4$ keV in the data taken during this observation. The ratio residuals in the bottom panel of Figure \ref{fig:2-10res} show this cut-off power law and broad Gaussian fit, with the parameter values presented as Model (2) Table \ref{tab:2-10params}. 

Figure \ref{fig:ironline} provides a focus on the $4-8$ keV energy range of the three data sets. Here a comparison is presented between a fit comprised of a cut-off power law fitting over the $2-10$ keV energy range excluding $4-7$ keV (i.e. to fit only the continuum) and Model (2) shown in Table \ref{tab:2-10params}. The difference in the width of the Fe K$\alpha$ emission line feature in the three observations is clearly shown here. The broad line in the \xmm data is in stark contrast with the very narrow feature found in the \suzaku data, suggesting interesting behaviour occurring in this already peculiar object (discussed in Section \ref{sect:discussion}). 

\subsection{Broadband X-ray Spectrum} 
\label{sect:broadband}
\begin{figure}
	\scalebox{1.0}{\includegraphics[width=\linewidth]{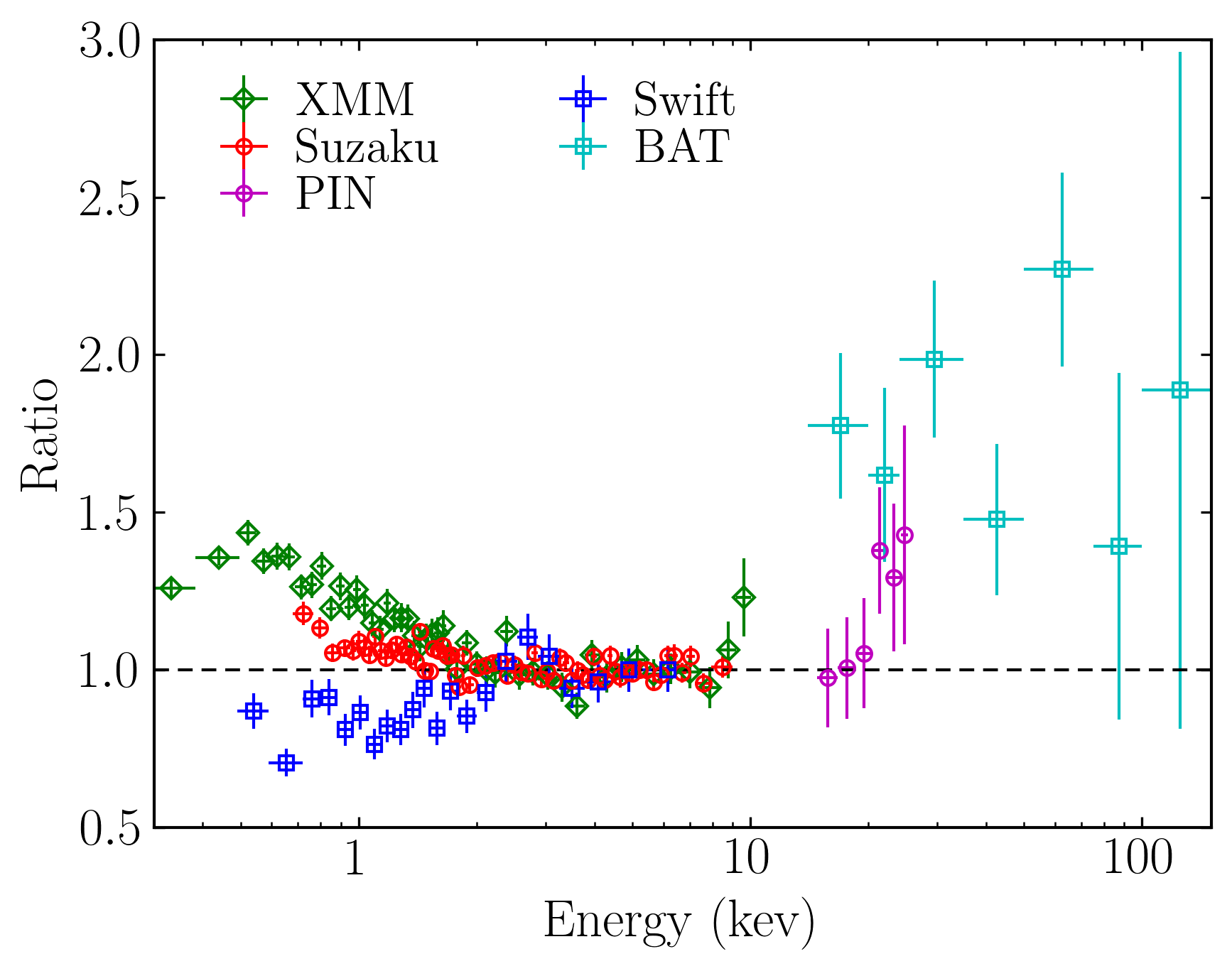}}
	\caption{The ratio residuals produced by extrapolating Model (2) from Table \ref{tab:2-10params} over the entire appropriate energy range for each data set. Clear low ($<2$ keV) and high ($>20$ keV) energy residuals are shown.}
	\label{fig:2-10extrap}
\end{figure}

Extrapolating Model (2) from Table \ref{tab:2-10params} to the broadband X-ray energy range appropriate for each data set produces the ratio residuals shown in Figure \ref{fig:2-10extrap}. The \xmm data display clear soft ($<2$ keV) residuals, suggesting the presence of a second component contributing to the spectrum. The \suzaku data do not show significant deviation from this extrapolated fit, though the data are truncated at $0.7$ keV, leaving out a large portion of where the soft excess would be. The PIN data above $\sim20$ keV are poorly fit, showing some curvature in the data toward higher energies. The model overestimates the \swift data below $\sim2$ keV with the BAT data being almost entirely underestimated, suggesting a more shallow broadband spectrum than found by Model (2) in the $2-10$ keV range.

\begin{figure}
	\scalebox{1.0}{\includegraphics[width=\linewidth]{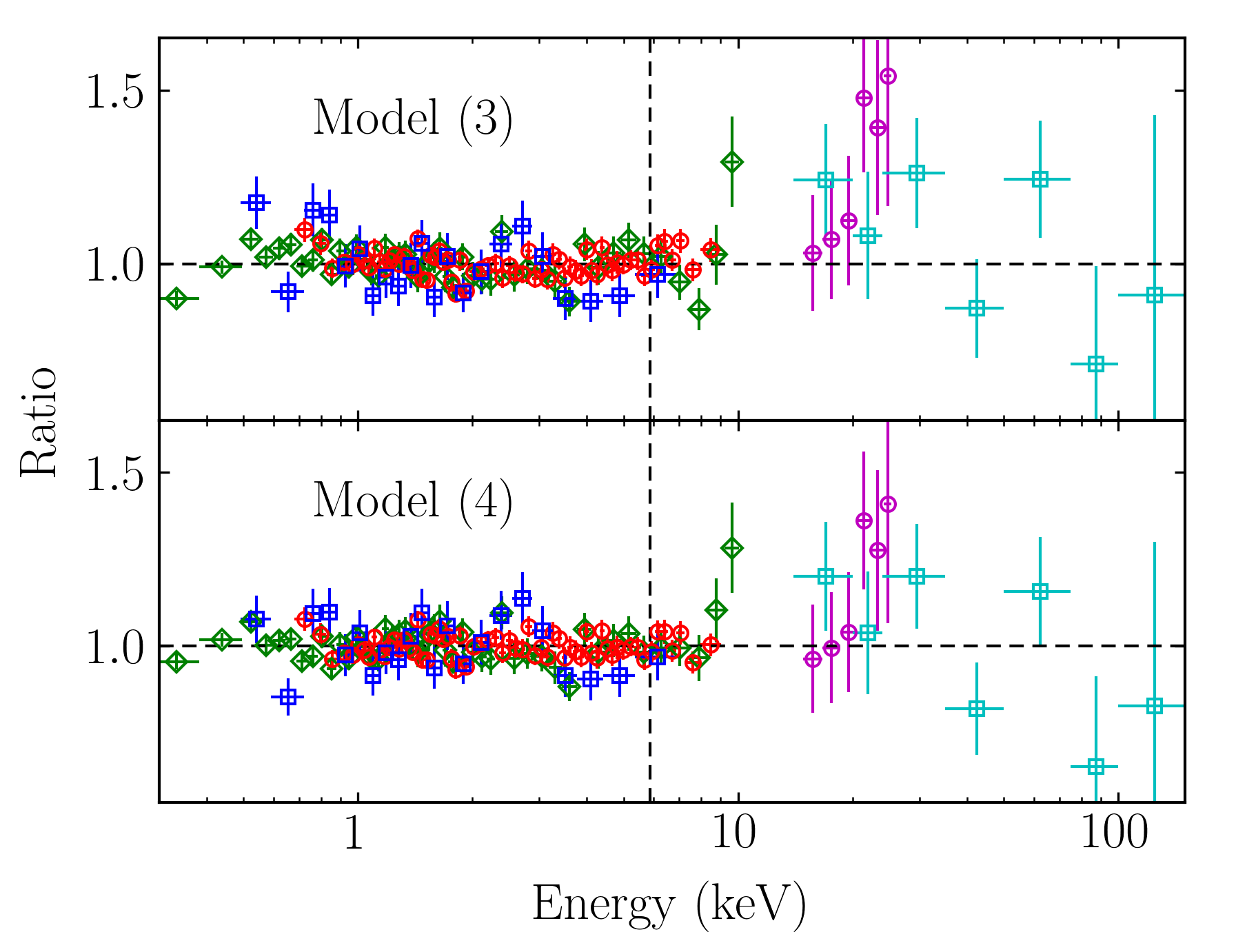}}
	\caption{The ratio residuals produced by Model (3), a cut-off power law and broad Gaussian, and Model (4), a cut-off power law, broad Gaussian, and black body,  presented in Table \ref{tab:simplebroadparams} fit over the appropriate energy ranges for each data set. Comparing the top and bottom panels shows similar fit quality, with the Model (4) displaying improved low energy residuals. Vertical lines in both the top and bottom panels denote the rest frame Fe K$\alpha$ line energy of $6.4$ keV.}
	\label{fig:simplebroadres}
\end{figure} 

%%% USING THE RBN SPECTRA
\begin{table*}
	\begin{center}
		\caption{The best fit parameters found by fitting the appropriate energy range for each of the three data sets of III~Zw~2. The \xmm broadband energy range is $0.3-10$ keV, \suzaku + PIN is $0.7-10$ keV and $15-25$ keV, and \swift + BAT is $0.5-7$ keV and $15-150$ keV. The normalisation of the power law component is in units of photons keV$^{-1}$ cm$^{-2}$ s$^{-1}$ at $1$ keV. The normalisation of the black body component is in units of $L_{39}/D^{2}_{10}$ where $L_{39}$ is the source luminosity in uints of $10^{39}$ erg s$^{-1}$ and $D_{10}$ is the distance to the source in units of $10$ kpc.}
		%\textbf{RBN SPECTRA C-STAT}
		\begin{tabular}{cccccccccc}                
			\hline
			Model & Data Set & $\Gamma$ & Norm$_{\mathrm{pl}}$ & $E_{\mathrm{K}\alpha}$ & $\sigma_{\mathrm{K}\alpha}$ & $EW_{\mathrm{K}\alpha}$ & $kT$ & Norm$_{\mathrm{bb}}$ & $\cstatdof$ \\
			& & & ($\times10^{-3}$) & (keV) & (keV) & (eV) & (eV) & ($\times10^{-6}$) & \\
			\hline
			\hline
			
			& \xmm & $1.80\pm0.02$ & $1.78\pm0.02$ & $6.67^{+0.74}_{-0.82}$ & $2.49^{+1.55}_{-0.81}$ & $1812^{+527}_{-695}$ & & & 160.02/131 \\
			(3) cpl+zgB & \suzaku + PIN & $1.58\pm0.01$ & $2.03\pm0.02$ & $6.41^{+0.13}_{-0.05}$ & $<0.215$ & $47^{+20}_{-25}$ & - & - & 192.67/156 \\
			& \swift + BAT & $1.57\pm0.02$ & $1.61\pm0.04$ & $6.28^{+0.18}_{-0.16}$ & $0.17^{+0.17}_{-0.11}$ & $251^{+169}_{-148}$ & & & 247.74/202 \\
			
			\hline
			
			& \xmm & $1.66^{+0.05}_{-0.04}$ & $1.59^{+0.07}_{-0.06}$ & $6.34^{+0.52}_{-0.47}$ & $1.02^{+1.01}_{-0.41}$ & $408^{+382}_{-354}$ & $173^{+13}_{-14}$ & $8.41^{+2.40}_{-2.59}$ & 138.69/129 \\
			(4) cpl+zgB+bb & \suzaku + PIN & $1.53\pm0.03$ & $1.89\pm0.07$ & $6.40^{+0.08}_{-0.05}$ & $<0.152$ & $38^{+20}_{-18}$ & $234^{+40}_{-51}$ & $4.93^{+1.94}_{-1.95}$ & 175.23/154 \\
			& \swift + BAT & $1.54\pm0.03$ & $1.52^{+0.07}_{-0.08}$ & $6.28^{+0.18}_{-0.16}$ & $0.18^{+0.17}_{-0.11}$ & $254^{+125}_{-146}$ & $144^{+56}_{-52}$ & $5.76^{+4.15}_{-3.20}$ & 238.85/200 \\
			
			\hline
			\label{tab:simplebroadparams}
		\end{tabular}
	\end{center}
\end{table*}

Refitting Model (2) to the broadband energy range of each data set vastly reduces the ratio residuals in all data sets, as shown in the top panel of Figure \ref{fig:simplebroadres}, with parameter values shown as Model (3) in Table \ref{tab:simplebroadparams}. The \xmm and \suzaku broadband fits are comparatively worse than the $2-10$ keV fits suggesting a more complex model is necessary while the \swift data is fit with approximately the same mediocrity. The width of the Gaussian feature in the \xmm fit has increased significantly, while the slope has simultaneously steepened, further indicating presence of a broad spectral feature and/or soft excess emission. 

Given the apparent soft excess in Figure \ref{fig:2-10extrap} the data were tested for the presence of a second spectral component by adding a black body component (denoted bb) to the previous model, producing the ratio residuals shown in the bottom panel of Figure \ref{fig:simplebroadres} and the model parameters shown as Model (4) in Table \ref{tab:simplebroadparams}. All three data sets show significant improvements to the quality of fit with the addition of a black body component to the model, though the \swift data are least affected by this additional spectral component. \xmm exhibits the steepest spectrum of the three, with the \suzaku and \swift fits showing agreement in photon index. 

Considering the time separation of 11$-$years between the \xmm and \suzaku observations, the fits in Table \ref{tab:simplebroadparams} can be interpreted as an evolution of the spectral shape of \obj due to the normalisation of the power law. During the \xmm observation, the lower power law normalisation would have exhibited a more significant low energy component (black body) while during the \suzaku observation a high power law normalisation would have acted to effectively over-power this second, weaker component. Further investigation of the spectrum of \obj requires two-component physically motivated models.

\begin{figure*}
	\scalebox{1.0}{\includegraphics[width=\linewidth]{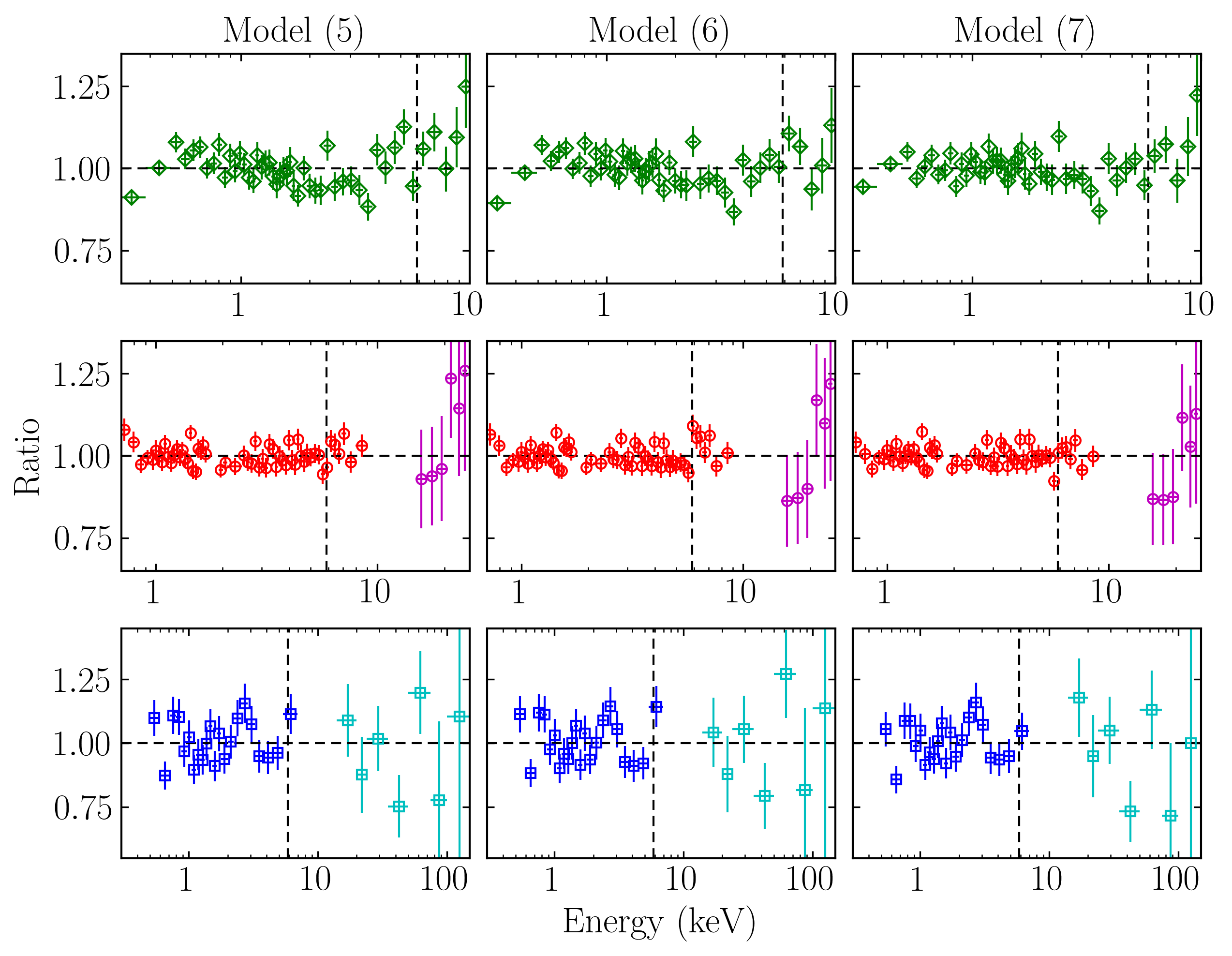}}
	\caption{The ratio residuals produced by the models presented in Table \ref{tab:xmmsuz_reflectionparams} fit over the appropriate energy ranges for each data set. Model (5) describes a cut-off power law and cold distant reflector, Model (6) a cut-off power law and blurred cold reflector, and Model (7) a cut-off power law and blurred ionised reflector. Comparing the three models shows very similar fit quality overall. Vertical lines in the panels denote the rest frame Fe K$\alpha$ line energy of $6.4$ keV.}
	\label{fig:individualbroadres}
\end{figure*}

%%% USING RBN SPECTRA
\begin{table*}
	\begin{center}
		\caption{Best fit model parameters for cold distant, Model (5), blurred cold, Model (6), and blurred ionised, Model (7), reflection.  Inclination angle is kept frozen at $i = 20^{\circ}$ in agreement with previous radio studies. The iron abundance was kept frozen at solar abundance $A_{\mathrm{Fe}} = 1.0$ and ionisation parameter was kept frozen at $\log\xi = 0.0$ erg cm s$^{-1}$ to best simulate a distant cold reflector (ie. the torus) in Models (5) \& (6). The iron abundance and ionisation parameter are left free to vary in Model (7). Subscript $p$ indicates that a parameter has been pegged at a limit.}
		%\textbf{RBN SPECTRA C-STAT}
		\begin{tabular}{cccccccccc}
			\hline
			Model & Data Set & $\Gamma$ & $\log F_{\mathrm{pl}}$ & $q$ & $A_{\mathrm{Fe}}$ & $\log\xi$ & $\log F_{\mathrm{ref}}$ & $R$ & $\cstatdof$ \\
			& & & (erg cm$^{-2}$ s$^{-1}$) & & & (erg cm s$^{-1}$) & (erg cm$^{-2}$ s$^{-1}$) & & \\
			\hline
			\hline
			
			& \xmm & $1.78\pm0.02$ & $-10.48\pm0.02$ & & & & $-10.99^{+0.10}_{-0.12}$ & $0.31^{+0.03}_{-0.04}$ & 181.17/133 \\
			(5) & \suzaku + PIN & $1.61\pm0.01$ & $-10.20\pm0.02$ & - & $1.0$ & $0.0$ & $-11.11^{+0.10}_{-0.13}$ & $0.12^{+0.01}_{-0.02}$ & 175.47/158 \\
			& \swift + BAT & $1.65\pm0.04$ & $-10.36\pm0.05$ & & & & $-11.08^{+0.14}_{-0.22}$ & $0.22^{+0.03}_{-0.05}$ & 241.39/204 \\
			
			\hline
			
			& \xmm & $1.81\pm0.02$ & $-10.53\pm0.02$ & & & & $-10.91^{+0.07}_{-0.09}$ & $0.42^{+0.03}_{-0.04}$ & 168.99/133 \\
			(6) & \suzaku + PIN & $1.64\pm0.02$ & $-10.24\pm0.03$ & $3$ & $1.0$ & $0.0$ & $-10.98^{+0.11}_{-0.16}$ & $0.18^{+0.02}_{-0.03}$ & 190.25/158 \\
			& \swift + BAT & $1.65\pm0.05$ & $-10.36\pm0.06$ & & & & $-11.10^{+0.17}_{-0.28}$ & $0.20^{+0.04}_{-0.06}$ & 245.93/204 \\
			
			\hline
			
			& \xmm & $1.64\pm0.02$ & $-10.40\pm0.02$ & $2.65^{+0.37}_{-0.38}$ & $4.78^{+3.04}_{-0.70}$ & $2.80^{+0.34}_{-0.08}$ & $-11.04^{+0.10}_{-0.12}$ & $0.23^{+0.02}_{-0.03}$ & 142.77/130 \\
			(7) & \suzaku + PIN & $1.62^{+0.02}_{-0.04}$ & $-10.21\pm0.02$ & $-10.00^{+10.37}_{p}$ & $0.51^{+0.25}_{-0.01p}$ & $1.40^{+0.62}_{-0.66}$ & $-10.94^{+0.10}_{-0.16}$ & $0.19^{+0.02}_{-0.03}$ & 170.68/155 \\
			& \swift + BAT & $1.63^{+0.08}_{-0.04}$ & $-10.33^{+0.04}_{-0.09}$ & $0.51^{+4.45}_{-10.51p}$ & $5.02^{+4.98p}_{-3.09}$ & $<2.48$ & $-11.15^{+0.14}_{-0.12}$ & $0.28^{+0.05}_{-0.04}$ & 237.18/201 \\
			
			\hline
			\label{tab:xmmsuz_reflectionparams}
		\end{tabular}
	\end{center}
\end{table*}

The results from phenomenological broadband fits above indicate a second component present in the spectra. To account for this physically, a reflector (e.g. the torus, accretion disc, etc.) illuminated by the hard X-ray source (corona or jet) can be modelled using the \texttt{XSPEC} model \texttt{xillver} \citep{Garcia+2014}.

A cold distant reflector (i.e. the torus) was modelled with \texttt{xillver} by setting the iron abundance to $A_{\mathrm{Fe}} = 1.0$ and the ionisation parameter to $\log\xi = 0.0$ erg cm s$^{-1}$. Results from superluminal motion of the radio jet in \obj require the inclination of the object to be $\leq41^{\circ}$ \citep{Brunthaler+2000}. With this we freeze the disc inclination at $i=20^{\circ}$ for all three data sets. Additionally, inclination was found to be poorly constrained when free to vary and does not improve the fit quality. The ratio residuals produced by the best fit model of a cold distant reflector are shown in the first column of Figure \ref{fig:individualbroadres} and the corresponding parameter values in Model (5) of Table \ref{tab:xmmsuz_reflectionparams}. 

The \suzaku fit is described by a bright power law that outshines a weaker reflection component. The \xmm fit shows a more steep but less prominent power law allowing more of the curvature intrinsic to the reflection spectrum to be seen in the average spectrum. Overall the data are fit well by this cold distant reflection model, though the broad lines found in the phenomenological models of \xmm and \swift are not typically the result of reflection from this type of material, prompting further modelling. Alternate models to \texttt{xillver}, like \texttt{reflionx} \citep{RossFabian2005} and \texttt{pexmon} \citep{Nandra+2007} were also attempted to model a cold reflector scenario. No one model seemed superior and we describe only the \texttt{xillver} fits here.

A blurred cold reflector was simulated by convolving the cold distant \texttt{xillver} reflection model with the blurring kernel \texttt{kdblur} (based off of \citealt{Laor1991} and first described by \citealt{Fabian+2002}). As in Model (5) the iron abundance and ionisation parameter are kept frozen at $A_{\mathrm{Fe}} = 1.0$ and $\log\xi = 0.0$ erg cm s$^{-1}$, respectively. The emissivity index in \texttt{kdblur} was kept frozen at $q = 3.0$ in accordance with a distant reflector. A maximally spinning Kerr black hole is assumed and as such the inner radius of the accretion disc was kept frozen at $R_{\mathrm{in}} = 1.235r_g$ and the outer radius of the disc was kept frozen at $R_{\mathrm{out}} = 400r_g$, where $r_g = GM/c^2$. The blurring inclination was linked to the inclination of the reflection component (i.e. frozen at $i=20^{\circ}$). Allowing these parameters to vary freely does not enhance the fit. The best fit model ratio residuals to a blurred cold reflector are displayed in the second column of Figure \ref{fig:individualbroadres} with the corresponding parameter values shown as Model (6) in Table \ref{tab:xmmsuz_reflectionparams}. 

The addition of a blurring kernel to the cold reflection model yields a significantly improved fit to the \xmm data while the \suzaku data are poorly fit. Conclusions about the physical system remain unchanged: the \suzaku data are best described by a bright power law that outshines the reflection component whereas the \xmm and \swift data exhibit a weaker power law component allowing curvature intrinsic to the reflection spectrum to be seen. We now explore the effect of iron abundance, ionisation parameter, and emissivity index on the fit quality to these data.

The broad lines observed in the spectra of the \xmm and \swift observations are indicative perhaps of blurred ionised reflection (i.e. off of the inner region of the accretion disc). Modelling such reflection with \texttt{xillver} convolved with \texttt{kdblur} requires allowing iron abundance, ionisation parameter, and emissivity index to be left free to vary. The best fit model of a blurred ionised reflector is presented in Model (7) of Table \ref{tab:xmmsuz_reflectionparams} with the ratio residuals shown in the bottom panel of Figure \ref{fig:individualbroadres}. 

Similar to the case of the previous models the \suzaku fit shows a strong power law component relative to reflection, with the \xmm and \swift fits finding slightly weaker power laws comparatively. Notably we find a poorly fit Fe K$\alpha$ line in the \suzaku data set with this blurred ionised model. We find that the broadened line of the \xmm data is best fit by this model, indicating the presence of relativistic line broadening due to emission originating from material in the accretion disc nearest the black hole \citep{Cunningham1975,Fabian+1989,Laor1991}. The reflection fraction ($R$) remains constant across the three data sets, suggesting the same amount of primary X-ray emission reaching the reflector in each observation. Most noteworthy are the extreme differences in iron abundance, ionisation parameter, and emissivity index between the \xmm and \suzaku data sets. The \xmm data is best fit by a super-solar abundance and highly ionised reflector whereas the \suzaku data correspond to a sub-solar abundance and less ionised reflector. Statistically this blurred ionised reflection model is the best fitting model for the \xmm and \swift data sets, though for the \suzaku data it offers no statistically significant improvement over the cold distant reflection scenario.

Comparing the results from Models (5), (6), and (7) we note that the \xmm data are better fit by reflection off of the inner accretion disc (Model 7) and that the \suzaku data are best fit by reflection off the torus (Model 5), prompting an interesting discussion as to the cause of this spectral difference over the course of only 11$-$years. 

Based on the best-fit model for each spectrum the observed $0.5-10$ keV flux is approximately $1.18\times10^{-11}$ ergs cm$^{-2}$ s$^{-1}$ and is comparable within $\sim15$ per cent between the epochs. The $10-100$ keV Swift BAT flux is approximately $2.48\times10^{-11}$ ergs cm$^{-2}$ s$^{-1}$. The intrinsic $2-10$ keV and $10-100$ keV luminosities are approximately $1.61\times10^{44}$ ergs s$^{-1}$ and $4.83\times10^{44}$ ergs s$^{-1}$, respectively, assuming a standard cosmology with $H_0 = 70$ km s$^{-1}$ Mpc$^{-1}$ and $\Omega_\Lambda = 0.73$.

\subsection{The X-ray Source} 
\label{sect:jetanalysis}
\begin{figure}
	\scalebox{1.0}{\includegraphics[width=\linewidth]{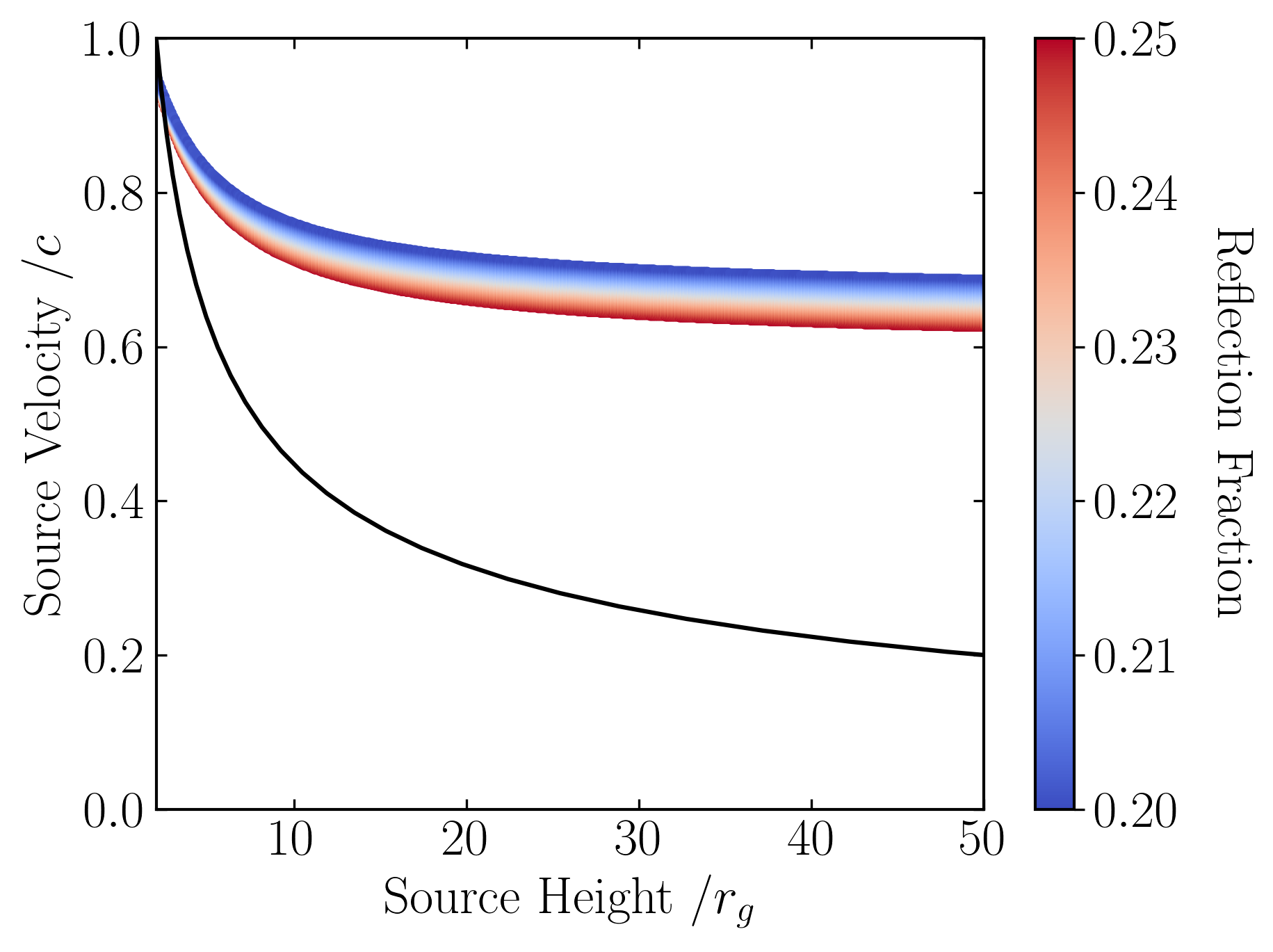}}
	\caption{The source height and source velocity parameter space given the reflection fraction of the best fit Model (7), a cut-off power law and blurred ionised reflector, to the \xmm data of \obj assuming a lamppost X-ray source. The black line represents the escape velocity of the black hole.}
	\label{fig:RZB}
\end{figure}
The \xmm data of \obj is found to be best fit by Model (7), a cut-off power law and blurred ionised reflector. This model describes relativistically broadened emission originating from a highly ionised, super-solar abundance accretion disc that is being illuminated by a primary X-ray source some height above the disc (i.e. a corona or jet). 

\cite{Gonzalez+2017} performed general relativistic (GR) ray tracing simulations studying the illumination pattern of the accretion disc (i.e. emissivity profile) for various corona geometries and velocities, constructing an approximation for the reflection fraction, $R$, produced by point source coronae (i.e. the lamppost model) given the source height and source velocity (see Section 5.1 therein). 

Assuming the lamppost model here we use this approximation of $R$ to compute the source height and source velocity parameter space that produces the reflection fraction value in Model (7) of $R=0.23^{+0.02}_{-0.03}$. These results are shown in Figure \ref{fig:RZB} as the shaded region with the black curve corresponding to the escape velocity from the black hole in \obj using the mass estimate $M_{\mathrm{BH}} = (1.84\pm0.27)\times10^8 M_{\odot}$ (\citealt{Grier+2012}). 

It is found that for all source heights $z > 2.4r_g$ the source velocity exceeds the escape velocity of the black hole. This results in an anisotropically emitting X-ray source moving away from the black hole and accretion disc system. A constraint on the source height cannot be made, however, for the source heights explored ($z = 2-50r_g$) the source velocity can be constrained as $v>0.6c$.

%####################%
\section{Discussion}
%####################%
\label{sect:discussion}
\obj is an interesting radio-intermediate Seyfert 1 galaxy. Its radio emission has been studied numerous times revealing extreme, possibly periodic, variability and the presence of a superluminal radio jet. The high-energy emission, however, has been studied far less with data being more sparse than at radio wavelengths. Here we have presented a thorough analysis of the X-ray spectrum of \obj obtained from the \textit{XMM-Newton}, \textit{Suzaku}, and \textit{Swift} satellites. 

All of the models presented throughout Section \ref{sect:meanspectrum} provide good statistical fits to the data sets analysed here. Model (3) of Table \ref{tab:simplebroadparams} was built upon by adding a black body emitter to produce Model (4), which significantly improved all of the fits, indicating a two-component spectrum. The addition of this second, soft component altered the slope of the power law ($\Gamma$) such that all three data sets were in agreement. Interestingly, it was found that the \xmm data were best fit by a broad ($\sigma_{\mathrm{K}\alpha} \approx 1$ keV) emission line at $E_{\mathrm{K}\alpha} \approx 6.3$ keV whereas the \suzaku data were best fit by a narrow ($\sigma_{\mathrm{K}\alpha} \approx 1$ eV) emission line at $E_{\mathrm{K}\alpha} \approx 6.4$ keV. The \swift data, however, are best explained as a mid-point between these two extremes, with parameter values resembling a blend of the other fits, which seems reasonable since \swift data are combined over 17$-$years. 

By simulating physically motivated models of reflection off of either the torus or accretion disc we were able to determine that the \xmm data were best fit by a blurred ionised reflector and that the \suzaku data were best fit by a cold distant reflector. Here again the best fit to the \swift data appears to be an amalgamation of these two very different fits, with parameter values residing between the lower and upper bounds provided by the \xmm and \suzaku fits. Combining the super-solar iron abundance and high ionisation parameter values of the \xmm fit suggests emission coming from an accretion disc illuminated by an X-ray emitting corona (e.g. \citealt{Fabian+2009,Fabian+2013,Gallo+2015}). The \suzaku fit, however, indicates a reflector with solar iron abundance that is not subject to the intense gravitational effects associated with being near the black hole as evidenced by the lower ionisation parameter. From this we infer that the \xmm data were taken during a period when the reflected X-ray emission of \obj was due to reflection off of the inner accretion disc whereas the \suzaku data were taken during a period of reflection off of the distant dusty torus surrounding the innermost region of the AGN, with the \swift data representing data taken during the transition between these two phases. The behaviour seems akin to that of changing-look AGN that appear to switch from Seyfert 1 to Seyfert 2 in the optical and X-ray (e.g. \citealt{LaMassa+2015}). However, unlike that class of AGN, \obj does not exhibit a marked change in the X-ray flux from 2001 to 2011. 

The most significant result in the work presented here is the discovery that a cold distant reflector provides the best fit to the \suzaku data, while a blurred ionised reflector produced the best fit to the \xmm data. This suggests that the source of the reflection completely changed over the course of only 11$-$years between these observations though the illuminating hard X-ray source has remained relatively constant as evidenced by the comparable flux. Considering only the \xmm and \suzaku data in combination with results from radio light curve analysis we hypothesise the cause of the strange behaviour exhibited by \obj is due to a precessing radio jet, as first mentioned by \cite{Brunthaler+2003,Brunthaler+2005}, being described in more detail later by \cite{Li+2010}. \cite{Brunthaler+2003,Brunthaler+2005} discuss a periodicity of approximately five years of the radio emission from \obj and mention a precessing jet as a potential explanation to the observations. In their description the jet would need to be interacting with a molecular torus on a five year cycle, though no evidence of a molecular torus was found. \cite{Li+2010} measured a quasi-periodic cycle in the radio light curve with period $P = 5.14 \pm 0.19$ years, supporting the approximation provided by \cite{Brunthaler+2003}. In addition, \cite{Clements+1995} were able to determine an approximate 13$-$month lag between X-ray and radio variability. 

The spectral differences observed between the two epochs can be explained as the radio jet precessing in such a way as to illuminate the inner disc region during the \xmm observation producing the broad Fe K$\alpha$ line and blurred ionised reflection spectrum, and then illuminating the torus during the \suzaku observation and producing signatures of a cold distant reflector in the X-ray spectrum. 

This evidence, while not conclusive, is supportive of a precessing radio jet interacting with a changing reflector producing the different X-ray spectra measured by \xmm in 2000 and \suzaku in 2011. Furthermore, the evidence of a tidal bridge with star forming knots between \obj and III~Zw~2B found by \cite{Surace+2001} is indicative of an ongoing merger phase which provides a plausible mechanism that could result in disc instabilities, in turn producing a precessing accretion disc that would influence the behaviour of the radio jet. 

Very recently, \cite{Liska+2017} showed via general relativistic magnetohydrodynamics (GRMHD) simulations that precessing accretion discs can in fact propagate radio jets along their spin axis, producing jets that precess with the disc. The interaction between \obj and III Zw 2B may have resulted in a precessing accretion disc, thus causing the radio jet to precess along with it, producing the spectral results presented here. 

The blurred reflection model for the \xmm data yield a low reflection fraction ($R \approx 0.23$) that is suggestive of a primary source that is anisotropically illuminating away from the accretion disc. This is plausible if the primary source is the base of the jet, which would be consistent with the radio jet in this source. Following the determination of $R$ from source height and source velocity in Section 5.1 of \cite{Gonzalez+2017} the measured $R$ value implies a source velocity of $v > 0.6c$ for source heights of $z = 2-50r_g$. This velocity is comparable to the radio jet velocity $v=0.874c$ (\citealt{Chen+2012}) supporting the connection between the X-ray source and radio jet. 

The observed spectral changes found here are perhaps reminiscent of the state transitions seen in X-ray binaries (XRBs). During epochs when the jet is on (i.e. the low/hard state) spectra from XRBs are described by emission from an optically thin, geometrically thick accretion disc. Once the jet has turned off (i.e. the high/soft state) the spectra from these objects are described by a traditional optically thick, geometrically thin standard accretion disc. Here we are unable to determine conclusively such a relationship between the radio jet and accretion disc as we lack simultaneous radio and X-ray observations. Additionally, the time scales for which such processes would occur in AGN are orders of magnitude greater than those of X-ray binaries and would therefore not be expected. 

Alternative interpretations of the data include disc truncation (e.g. \citealt{Lohfink+2013}) and ionisation changes in the disc (e.g. \citealt{Bonson+2015,Gallo+2015}). In a disc truncation scenario the inner portion of the accretion disc would be lost due to launching, ejecting the material via a jet. This, however, would almost certainly cause flux changes as a portion of the reflector is lost. With the essentially constant flux between the \xmm and \suzaku observation it is unlikely that a disc truncation scenario would produce the observed spectral differences. 

Changes in the ionisation of the accretion disc would be associated with changes in the source flux. If the disc is exposed to a primary source that is brightening one may see a correlation between the flux and ionisation parameter. However, if light bending of the primary source is important (e.g.\citealt{MiniuttiFabian2004}) then a correlation between brightness and ionisation may not be obvious. Nevertheless, the flux during the \xmm and \suzaku observations was comparable and unlikely to result in significant changes in the ionisation that would alter the appearance of the broad emission line.

Alternatively, increased Comptonisation due to a variable, patchy corona could alter the shape of the reflection spectrum (i.e. broad line) (e.g. \citealt{WilkinsGallo2015a}). If a patchy corona were covering a large fraction of the central disc then the blurred emission would be further Comptonised as it traverses the corona. In this scenario, the corona could  be more obscuring during the Suzaku observation and hence Comptonise the broad line that is evident during the XMM observation. This may be consistent with the slightly high power law flux observed in 2011.

%#####################%
\section{Conclusions}
%#####################%
\label{sect:conclusions}
\obj is a peculiar radio-intermediate Seyfert 1 galaxy that exhibits extreme variability in its radio emission. The most recent set of X-ray observations have been used to perform a spectral analysis on the X-ray spectrum, which has been studied far less than the radio emission.

We find evidence of a soft excess in all data sets with a broad Fe K$\alpha$ line required in the \xmm and \swift data, but only a narrow line in the \suzaku data. 

Physical models of a cold distant reflector (i.e. the torus) and blurred ionised reflector (i.e. the inner region of the accretion disc) were fit to the data, finding the \xmm data best described by a blurred ionised reflector (i.e. the accretion disc) and the \suzaku data being best fit by a cold distant reflector (i.e. the torus).

The spectral difference observed in the \xmm and \suzaku data sets, which occurred over the 11$-$years between observations, is hypothesized as the result of a precessing radio jet illuminating the inner disc region in 2000 and interacting with the dusty torus in 2011. 

However, other possible explanations exist. Continued monitoring in the radio and X-ray wavelengths as well as deep pointed observations with \xmm and \nustar will help better understand the nature of the X-ray variability. 

%###########################%
\section*{Acknowledgements}
%###########################%
This work made use of data supplied by the UK Swift Science Data Centre at the University of Leicester. We thank the anonymous referee for their careful reading and helpful comments on the original manuscript.

%%%%%%%%%%%%%%%%%%%%%%%%%%%%%%%%%%%%%%%%%%%%%%%%%%


\bibliographystyle{mnras}
\bibliography{IIIZw2_refs} % if your bibtex file is called example.bib

% Don't change these lines
\bsp	% typesetting comment
\label{lastpage}
\end{document}